\documentclass[a4paper,11pt]{article}
\usepackage{a4}
\usepackage{parskip} 
\usepackage{graphicx}    
\usepackage{subfigure}
\usepackage{amsmath, amsthm, amssymb}
\usepackage{texab}
\usepackage{sfspaper}
\usepackage{natbib}

\begin{document}

\newtheorem{theorem}{Theorem}[section]
\newtheorem{proposition}[theorem]{Proposition}

\newtheorem{definition}{Definition}[section]

\newcommand{\na}{\texttt{na}} \newtheorem{algo}{Algorithm}[section]

\bibliographystyle{apalike}

\Title{Rate estimation in partially observed Markov jump processes with
  measurement errors}

\Author{Michael Amrein and Hans R. K\"unsch}

\maketitle

\begin{abstract}\normalsize

  We present a simulation methodology for Bayesian estimation of rate
  parameters in Markov jump processes arising for example in stochastic
  kinetic models. To handle the problem of missing components and
  measurement errors in observed data, we embed the Markov jump process
  into the framework of a general state space model. We do not use
  diffusion approximations. Markov chain Monte Carlo and particle filter
  type algorithms are introduced, which allow sampling from the posterior
  distribution of the rate parameters and the Markov jump process also in
  data-poor scenarios. The algorithms are illustrated by applying them to
  rate estimation in a model for prokaryotic auto-regulation and in the
  stochastic Oregonator, respectively.\\

\end{abstract}

\section{Introduction}
It is generally accepted that many important intracellular processes, e.g.
gene transcription and translation, are intrinsically stochastic, because
chemical reactions occur at discrete times as results from random molecular
collisions (\cite{ma} and \cite{arm}). These stochastic kinetic models
correspond to a Markov jump process and can thus be simulated using techniques
such as the Gillespie algorithm (\cite{g}) or -- in the time-inhomogeneous 
case -- Lewis' thinning method
(\cite{ogata}). Many of the parameters in such models are uncertain or
unknown, therefore one wants to estimate them from times series data. 
One possible approach is to approximate the model with a diffusion and then
to perform Bayesian (static or sequential) inference based on the
approximation (see \cite{gw}, \cite{gwII}, \cite{gwIII} and
\cite{gwbookart}). This gives more flexibility to generate the proposals
(see \cite{dg}), but it is difficult to to quantify the approximation error. Depending on the
application, it might be preferable to work with the original Markov
jump process. This possibility is mentioned in \cite{wil}, cahpter 10, and \cite{bwka}
demonstrate in the case of the simple 
Lotka-Volterra model that this approach is feasible in principle, but
in more complex situations it is difficult to construct a Markov chain
Monte Carlo (MCMC) sampler with good mixing properties. The key problems
in our view are to construct good proposals for the latent process
on an interval when the values at the two end points are fixed and
the process is close to the boundary of the state space, and to construct
reasonable starting 
values for the process and the parameters, in particular when some of
the components are observed with small or zero noise. We propose here solutions
for both of these problems that go beyond \cite{wil}, Chapter 10, and \cite{bwka} and thus
substantially enlarge the class of models that are computationally
tractable.

The rest of the paper is organized as follows. In Section \ref{sd}, we
describe the model, establish the relation to stochastic kinetics and
introduce useful notation and densities. In Section \ref{bamcmc}, we
motivate the Bayesian approach and present the base frame of the MCMC
algorithm. Section \ref{gen} describes in detail certain aspects of the
algorithm, mainly the construction of proposals for the latent Markov jump
process. In Section \ref{ini}, the particle filter type algorithm to
initialize values for the parameters and for the latent Markov jump process is
presented. In Section \ref{ex}, we look at two examples. First, the
stochastic Oregonator (see \cite{g}) is treated in various scenarios,
including some data-poor ones, to show how the algorithm works. Then, we
turn to a model for prokaryotic auto-regulation introduced in \cite{gw} and
reconsidered in \cite{gwbookart}. Finally, conclusions are given in Section
\ref{concl}.

\section{Setting and definitions}\label{sd}
\subsection{Model}\label{model}
Consider a Markov jump process $\mathcal{Y}=\{y_t=(y^1_t,\dots,y^p_t)^T:t
\geq t_0 \}$ on a state space $\mathcal{E} \subset 
\mathbb{N}_0^p$ with jump vectors $A_i \in \mathbb{Z}^p$ for $i \in \{1, \dots
,r\}$ and possibly time dependent transition intensities
$\mu_i(t,y)=\theta_i\cdot h_i(t,y) $:  
\begin{equation*}
\textrm{P}[y_{t+\delta}=y+A_i|y_t=y]=\mu_i(t,y)\delta+o(\delta) \ \
(\delta>0).
\end{equation*}
We denote the total transition intensity by
$$ \mu_0(t,y) = \sum_{i=1}^r \mu_i(t,y).$$
We assume that the functions $h=\{h_i\}_{i \in \{1,2, \dots ,r\}}$, called
the standardized transition intensities, the jump matrix $A$ with
columns $A_i$ and the initial distribution $f_0$ of $y_{t_0}$ are
known. The goal is to estimate the hazard rates 
$\theta=(\theta_1,\dots,\theta_r)$ from partial measurements
$x_{0},x_{1},\dots,x_{n}$ of the process
at discrete time points $0=t_0<t_1<\dots<t_n$. Unobserved components are
set as $\na$ and we assume

$$x_l|\mathcal{Y}=x_l|y_{t_l}\sim g_{\eta}(.|y_{t_l}),$$ 

where $g_{\eta}(x_l|y_{t_l})$ 
is a density with respect to some $\sigma$-finite measure
 (with possibly unknown) nuisance parameter $\eta$.
We specify this more precisely in the examples in Section \ref{ex}.

This framework can be regarded as a general state space model: $x_0,x_1,\dots$
is an observed times series which is derived from the unobservable Markov
chain $y_{t_0},y_{t_1},\dots$ (see \cite{kue} or \cite{dou}). 

For computational reasons, we further assume that we can easily evaluate
the time-integrated standardized transition intensities
\begin{equation*}
H_i(s,t,y):=\int_{s}^t h_i(u,y) du.
\end{equation*}

Models of the above form arise for example in the context of stochastic
kinetics. Consider a biochemical reaction network with $r$ reactions
$R_1,\dots, R_r$ and $p$ species $Y^1,\dots,Y^p$, i.e.,
\begin{equation*}
R_i: \ v_{i1}Y^1+v_{i2}Y^2+\dots +v_{ip}Y^p \ 
\longrightarrow \  u_{i1}Y^1+u_{i2}Y^2+\dots +u_{ip}Y^p
\end{equation*}
for $i=1,\dots ,r$. Let $y^j_t$ denote the number of species $Y^j$
at time $t$, $y_t=(y^1_t,\dots,y^p_t)^T$, $V=(v_{ij})$ and
$U=(u_{ij})$. Then, according to the mass action law, we can describe
$\{y_t:t\geq t_0\}$ as a Markov jump process with jump matrix
$A=(U-V)^T$ and standardized reaction intensities
\begin{equation*}
h_i(y)= \prod_{j,v_{ij}\geq 1}
\left({y^j\atop~v_{ij}}\right).
\end{equation*}
For further details, see e.g. \cite{g}, \cite{gw} or \cite{gwbookart}.
We will use in the following terminology from this application: We 
will call the jump times reaction times and classify a jump as one
of the $r$ possible reaction types. 

\subsection{ Additional notation and formulae for densities}
A possible path $y_{[a,b]}$ on an interval $[a,b]$ in our model is uniquely
characterized by the total number of reactions $n_{tot}$, the
initial state $y_a$, the successive reaction times $a < \tau_1 < \ldots
< \tau_{n_{tot}} \leq b$ and the reaction types (or indices)
$r_1, r_2, \ldots, r_{n_{tot}} \in \{1, \ldots, r\}$. The states at the
reaction times are then obtained as
$$y_{\tau_k} = y_a + \sum_{i=1}^k A_{r_i}.$$
We write for simplicity $y_k$ instead of $y_{\tau_k}$. 
Furthermore, $r_{tot}^i$ is the total number of reactions of type $i$
and $r_{tot}$ is the vector with components $r_{tot}^i$. All these
quantities depend on the interval $[a,b]$. If this interval is
not clear from the context, we write $n_{tot}([a,b])$, $\tau_k([a,b])$, etc.

The density $\psi_{\theta}$ of $y_{[a,b]}$ given $y_a$ is well known, see
e.g. \cite{wil}, Chapter 10. Defining $\tau_0=a$, $\tau_{n_{tot}+1}=b$ and $y_0=y_a$, it is given by
\begin{equation*} 
\begin{split}
\psi_{\theta}(y_{[a,b]}|y_a)
&=\exp \left(-\sum_{i=1}^r\theta_{i}
    \int_{a}^{b}h_i(s,y_s)ds \right)\cdot
  \prod_{k=1}^{n_{tot}} \theta_{r_k}h_{r_k}(\tau_k,y_{{k-1}}) \\
&=\exp \left(-\sum_{k=1}^{n_{tot}+1} \sum_{i=1}^r\theta_{i}
    H_i(\tau_{k-1},\tau_{k},y_{{k-1}})\right) \cdot \prod_{k=1}^{n_{tot}}\theta_{r_k}h_{r_k}(\tau_k,y_{{k-1}}).
\end{split}
\end{equation*}
In the time-homogeneous case, i.e., $h_i(t,y)=h_i(y)$, we have 
$H_i(\tau_{k-1},\tau_{k},y_{{k-1}})=h_i(y_{{k-1}})\delta_k$
with $\delta_{k}=\tau_k-\tau_{k-1}$. Therefore
\begin{equation}\label{gill1}
\delta_k|\tau_{k-1},y_{k-1} \sim \textrm{Exp}(\mu_0(y_{k-1}))
\end{equation}
and 
\begin{equation}\label{gill2}
  \textrm{P}[r_k=i|\tau_{k-1},y_{k-1}]=\frac{\mu_i(y_{k-1})}{\mu_0(y_{k-1})},
\end{equation}
and we can exactly simulate the Markov jump process using the Gillespie
algorithm (see 
\cite{g}) or some faster versions thereof (see \cite{gb}). Replacing
$h_i(y_{k-1})$ by $h_i(\tau_{k-1},y_{k-1})$ in 
(\ref{gill1})  and (\ref{gill2}), this can be done ``approximately'' in the
inhomogeneous case. An exact simulation algorithm based on a thinning
method is described in \cite{ogata}. 

We write the density of all observations in $[a,b]$ as
\begin{equation*}
g_{\eta}(x_{[a,b]}|y_{[a,b]})=\prod_{l,a \leq t_l
\leq b}g_{\eta}(x_l|y_{t_l}),
\end{equation*}
where the empty product
is interpreted as $1$. The joint density of $y_{[t_0,t_n]}$ and
$x_{[t_0,t_n]}$ (given the
parameters $\theta$ and $\eta$) is then
\begin{equation} \label{denscompl}
p(y_{[t_0,t_n]},x_{[t_0,t_n]}|\theta,\eta)
=f_0(y_{0})\cdot \psi_{\theta}(y_{[t_0,t_n]}|y_0)\cdot
g_{\eta}(x_{[t_0,t_n]}|y_{[t_0,t_n]}).
\end{equation}

\section{Bayesian approach and Monte Carlo methods}\label{bamcmc}

The maximum likelihood estimator 
is too complicated to compute because we are not able to calculate the
marginalisation of the density in (\ref{denscompl}) over
$y_{[t_0,t_n]}$ explicitly. It seems easier to combine a 
Bayesian approach with Monte Carlo methods, that is we will
sample from the posterior distribution of the parameters and the
underlying Markov jump 
process $y_{[t_0,t_n]}$ given the data (see \cite{rob}). This has also the
additional 
advantage that prior knowledge about the reaction rates can be used.
Assuming $\theta$ and $\eta$ to be independent a priori, the joint
distribution of $y_{[t_0,t_n]}$, $x_{[t_0,t_n]}$,
$\theta$ and $\eta$ has the form 
\begin{equation*}
p(y_{[t_0,t_n]},x_{[t_0,t_n]},\theta,\eta)=
p(y_{[t_0,t_n]},x_{[t_0,t_n]}|\theta,\eta)\cdot p(\theta) \cdot p(\eta).
\end{equation*}

We want to simulate from $p(y_{[t_0,t_n]},\theta,\eta|x_{[t_0,t_n]})$,
which also yields samples from $p(\theta,\eta|x_{[t_0,t_n]})$ using a
marginalisation over $y_{[t_0,t_n]}$. The standard approach to do this is
iterating between blockwise updates of the latent process 
$y_{[t_0,t_n]}$ on sub intervals of $[t_0,t_n]$ with
Metropolis-Hastings steps, updates of $\theta$  and
updates of $\eta$  (see e.g. \cite{gilks}, chapter 1,
\cite{bwka} or \cite{gwbookart}). 

As in \cite{bwka}, we choose independent Gamma distributions with
parameters $\alpha_i$ and $\beta_i$ as priors for $\theta_i$:
\begin{equation*}
p(\theta) \propto \prod_{i=1}^r\theta_{i}^{\alpha_i-1}\exp(-\beta_i
\theta_i) 
\end{equation*} 
We write this distribution as $\Gamma_r(\alpha,\beta)$ where 
$\alpha$ and $\beta$ are vectors of dimension $r$. 
Conditionally on $y_{[t_0,t_n]}$,$x_{[t_0,t_n]}$ and $\eta$,
the components $\theta_i$ have then again independent Gamma
distributions, more precisely
\begin{equation}\label{simtheta}
\theta|y_{[t_0,t_n]},x_{[t_0,t_n]},\eta \sim
\theta|y_{[t_0,t_n]} \sim
\Gamma_r\left(\tilde{\alpha}(y_{[t_0,t_n]}),\tilde{\beta}(y_{[t_0,t_n]})\right), 
\end{equation}
with 
\begin{equation*}
\tilde{\alpha}_i(y_{[t_0,t_n]},\alpha_i)=\alpha_i+r_{tot}^i
\end{equation*}
and
\begin{equation*}\tilde{\beta}_i(y_{[t_0,t_n]},\beta_i)=\beta_i+
\int_{t_0}^{t_n}h_i(s,y_s)ds=\beta_i+\sum_{k=1}^{n_{tot}+1}
H_i(\tau_{k-1},\tau_{k},y_{{k-1}}). 
\end{equation*}

Choosing a suitable prior for $\eta$ depends heavily on the error
distribution, so we 
refer to the examples in Section \ref{ex}.

We propose the following algorithm, which will be explained in more
detail in the next sections. The generation of initial values
$y^{(0)}_{[t_0,t_n]}$, 
$\theta^{(0)}$ and $\eta^{(0)}$ will be discussed in Section
\ref{ini}. The choice of the set $\mathcal{I}_{[t_0,t_n]}$ of 
overlapping subintervals $[a,b] \subset [t_0,t_n]$ for updating
$y$ will be discussed in Section \ref{sub-int}.

\begin{algo}[Simulation from $y_{[t_0,t_n]},\theta,\eta$ given
  $x_{[t_0,t_n]}$]\label{simyhetaeta}
\mbox{}
For $m=1,2,\dots,M$:
\begin{itemize}
\item[1.] Set
  $y_{[t_0,t_n]}=y^{(m-1)}_{[t_0,t_n]}$, $\theta=\theta^{(m-1)}$,
  $\eta=\eta^{(m-1)}$.
  Update $y_{[a,b]}$ for all $[a,b] \in \mathcal{I}_{[t_0,t_n]}$
  sequentially in fixed order by 
  proposing $y^{new}_{[a,b]}$ as described in sections \ref{genr},
  \ref{geny} and \ref{updateborder} and replacing  
  $y_{[a,b]}$ by $y^{new}_{[a,b]}$ with probability
  $\alpha(y^{new}_{[a,b]}|y_{[a,b]},\theta,\eta)$ (see
  (\ref{accprob})). 
 Set $y^{(m)}_{[t_0,t_n]}=y_{[t_0,t_n]}$. 
\item[2.] Simulate $\theta^{(m)} \sim
\Gamma_r\left(\tilde{\alpha}(y^{(m)}_{[t_0,t_n]}),\tilde{\beta}(y^{(m)}_{[t_0,t_n]})\right)$.
\item[3.] Generate $\eta^{(m)}$ given $y^{(m)}_{[t_0,t_n]}$ in a suitable fashion.
\end{itemize}
\end{algo}

\section{Simulating a path given parameters and 
observations}\label{gen}

We assume now that $\theta$ and $\eta$ are fixed and we want to modify 
$y_{[a,b]}$ on sub intervals $[a,b]$ of
$[t_0,t_n]$. First we consider the case $t_0 <a < b <t_n$ where
the values $y_a$ and $y_b$ remain unchanged. The boundary cases will
be discussed in \ref{updateborder}. Exact
methods to simulate from a continuous time Markov chain conditioned on both
endpoints are reviewed and discussed in 
\cite{hs}. The rejection method
is too slow in our examples, and the other two require eigendecompositions
of the generator matrix. This would require truncating the state space
and is too time-consuming in our examples. Hence we use a
Metropolis-Hastings procedure.  Our proposal distribution $q$ first 
generates a vector of new total reaction numbers $r_{tot}^{new}$ on $[a,b]$
and then, conditioned on $r_{tot}^{new}$, generates a value $y^{new}_{[a,b]}$. 

\subsection{Generating new reaction totals}\label{genr}
Because the values
$y_a$ and $y_b$ are fixed, we must have that
\begin{equation}
\label{condrtot} 
A r_{tot}^{new}=y_b-y_a=A r_{tot} \Leftrightarrow 
A(r_{tot}^{new}-r_{tot})=0  .
\end{equation}
If $rank(A)=r$, the reaction totals remain unchanged. Otherwise
it is known that $\{x \in \mathbb{Z}^r: A\cdot x=0\}$ forms a
lattice and can be written as $\{a_1\cdot v_1+\dots + a_d\cdot
v_d:a_1,\dots,a_d \in \mathbb{Z}\}$ with  $d=\dim(\ker(A))$ and
basis vectors $v_l\in\mathbb{Z}^r $, $l \in \{1,2,\dots,d\}$ (note that
these vectors are not unique). Appendix \ref{dioeq} describes how to
compute a basis vector matrix
\begin{equation*}
V(A)=(v_1,\dots,v_d).
\end{equation*}
This enables us to generate
a vector $r_{tot}^{new}$ which respects (\ref{condrtot}) in
a simple way: 
\begin{equation}\label{simrtotnew}
r_{tot}^{new}=r_{tot} +V(A)\cdot Z
,\ Z \sim q_{\iota}^Z,
\end{equation}
where $q_{\iota}^Z$ is a symmetric proposal distribution $q_{\iota}^Z$ on
$\mathbb{Z}^d$, i.e.,  $q_{\iota}^Z(z)=q_{\iota}^Z(-z)$, with parameter
$\iota$. If
$r_{tot}^{new}$ has a negative component, we stop and 
set $y^{new}_{[a,b]}=y_{[a,b]}$. 

\subsection{Generating a new path given the reaction totals}
\label{geny}

The new path $y_{[a,b]}$ depends only on $y_a$ and the new reaction totals
$r^{new}_{tot}$, and not on the old path $y_{[a,b]}$.
The constraint $y^{new}_b=y_b$ is satisfied automatically
by our construction of $r^{new}_{tot}$. Therefore our algorithm 
simply generates a path on $[a,b]$ with given initial value and given
reaction totals, and we can omit the superscripts $new$.
\cite{bwka} generate the path 
according to $r$ independent inhomogeneous Poisson processes with intensities
\begin{equation*}
\lambda_i(t)=\mu_i(a,y_a)\frac{b-t}{b-a}+\mu_i(b,y_b)\frac{t-a}{b-a},
\end{equation*}
conditioned on the totals $r^{new,i}_{tot}$. In situations where the
standardized reaction intensities 
$h_i$ depend strongly on $y$, this proposal often generates
paths that are impossible under  
the model. This is typically the case when the
number of molecules of some species is small.
Our proposal first 
 decides the order in which the reactions take place,
that is we first generate $r_k$ for $k=1,2, \ldots, n_{tot}$. In a
second step, we generate the reaction times $\tau_k$, taking into account
both the probability of a reaction of a given
type at the current state of the process and the remaining number of
reactions $S_k^i$ of type $i$  
after time $\tau_k$ that still have to occur in order to reach the 
prescribed total.  In order to 
make the description of the algorithm easier to read, we mention that $t_k^*$ is a first
guess for $\tau_{k-1}$ (needed only if the intensities are time 
inhomogeneous). Also remember that $y_k=y_{\tau_k}$. 

\begin{algo}[Generating $y_{[a,b]}$ given
  $r_{tot}$ and $y_a$]\label{propyb}
\mbox{}
\begin{enumerate}
\item Set 
  $S_0^i =r_{tot}^i$ for $i \in \{1,\dots,r\}$ and $y_0=y_a$.
\item For $k=1,\dots,n_{tot}$ do the following:\\
  Set $t^*_{k}=a + (b-a)(k-1)/n_{tot}$.
  If $\mu_l(t^*_k,y_{k-1})=0$ for all $l$ with $S^l_{k-1}>0$,
  stop. Otherwise, generate $r_k$ with probabilities
  \begin{equation}\label{probspropy}
\mathrm{P}[r_k=i]\propto \sqrt{S^i_{k-1} \mu_i(t^*_k,y_{k-1})}.
  \end{equation}
  If $r_k=i$, set 
  $S^i_{k} = S^i_{k-1}-1$, $S^l_{k}=S^l_{k-1}$
  for $l \neq i$ and $y_{k}=y_{k-1}+A_i$. 
\item Generate $(\delta_k; k\in\{1,\dots,n_{tot}+1\})$
  according to a Dirichlet distribution with parameter $\alpha =(\alpha_k; 
   k\in\{1,\dots,n_{tot}+1\})$ where 
\begin{equation} \label{alphadir}
\alpha_k= \mu^{-1}_0(t_k^*,y_{k-1})\frac{\sum_{l}\mu^{-1}_0(t_l^*,y_{l-1})}
{\sum_{l} \mu^{-2}_0(t_l^*,y_{l-1})},
\end{equation}
and set $\tau_k=\tau_{k-1} + (b-a)\delta_k$ 
for $k=1,\dots,n_{tot}$.
\end{enumerate}
\end{algo}

The algorithm stops in step 2 when we can no longer reach the state $y_b$
on a possible reaction path using the available remaining reactions.
This means that an impossible path is proposed which has acceptance
probability 0. 

The heuristics behind the steps in the above algorithm is the following.
The probabilities (\ref{probspropy}) are an attempt to reach
a compromise between the probability of a reaction of type $i$ at the
current state according to the law of the process 
and the remaining number of reactions of type $i$ that
still have to occur in order to reach the prescribed total. 
 Empirically, we found that choosing these probabilities  
proportional to the the geometric mean leads to good acceptance
rates in the examples in Section \ref{ex}. The Dirichlet distribution
in (\ref{alphadir}) is used as an approximation of the
distribution of independent exponential-($\mu_0(t_k^*,y_{k-1})$) waiting
times $\delta_k$ conditioned on the event that their sum is equal to $b-a$.
If all $\mu_0(t_k^*,y_{k-1})$ are equal, the conditional first two
moments are 
\begin{equation}
\label{Dirichlet-mean}
\mathrm{E}\left[ \delta_k \mid \sum_l \delta_l = b-a\right] = (b-a) \frac{\mathrm{E}\left[\delta_k\right]}
{\sum_l \mathrm{E}\left[\delta_l\right]}
\end{equation}
and
\begin{equation}
\label{Dirichlet-var}
\mathrm{Var}\left[\delta_k\mid\sum_l \delta_l=b-a\right] = (b-a)^2\left(\frac{\mathrm{Var}(\delta_k)+\mathrm{E}[\delta_k]^2}{\sum_l\mathrm{Var}(\delta_l)+(\sum_l\mathrm{E}[\delta_l])^2}-\left(\frac{\mathrm{E}\left[\delta_k\right]}
{\sum_l \mathrm{E}\left[\delta_l\right]}\right)^2\right) ,
\end{equation}
and moreover the conditional distribution
is Dirichlet with parameters $\alpha_k=1$, scaled by $b-a$, see e.g. \cite{BicPD77}, Section 1.2. 
In the general case, we use a Dirichlet distribution as approximation
and determine
the parameters such that the expectation matches the right-hand 
side of (\ref{Dirichlet-mean}) for all $k$. This implies that
$\alpha_k \propto \mu^{-1}_0(t_k^*,y_{k-1})$. Finally, the proportionality 
factor is determined such that 
the sum of the variances
matches the sum of the right-hand side of (\ref{Dirichlet-var}).

\subsection{Acceptance probability of a new path}
By construction, the proposal density $q(y^{new}_{[a,b]}|y_{[a,b]},\theta)$ 
has the form
$$q(y^{new}_{[a,b]}|y_a,r^{new}_{tot},\theta) q(r^{new}_{tot}|r_{tot})$$
Because of the symmetry of $q_{\iota}^Z$, we have
\begin{equation*}
q(r_{tot}^{new} | r_{tot})=q(r_{tot} |r_{tot}^{new}).
\end{equation*}
So it will cancel out in the
acceptance probability and we do not need to consider it.

Next, $q(y_{[a,b]}|y_a,r_{tot},\theta)$ is equal to 
\begin{equation*}
\prod_{k=1}^{n_{tot}}
\frac{\sqrt{S^i_{k-1} \mu_i(t_k^*,y_{k-1})}}{\sum_{l=1}^{r}\sqrt{S^l_{k-1}
    \mu_l(t_k^*,y_{k-1})}}\cdot
\frac{f^{\mathrm{Dir}}_{\alpha}\left((\tau_k-\tau_{k-1})/(b-a):k\in\{1,\dots,n_{tot}+1\}\right)}{(b-a)^{n_{tot}}}
\end{equation*}
where $f^{\mathrm{Dir}}_{\alpha}$ is the density of the Dirichlet distribution with
parameter $\alpha$ from (\ref{alphadir}).

Hence, according to the Metropolis-Hastings recipe, the 
acceptance probability is
\begin{equation}\label{accprob}
\alpha(y^{new}_{[a,b]}|y_{[a,b]},\theta,\eta)
=\min\left\{1, \frac{\psi_{\theta}(y^{new}_{[a,b]}|y_a)
g_{\eta}(x_{[a,b]}|y^{new}_{[a,b]}) q(y_{[a,b]}|y_a,r_{tot},\theta)} 
  {\psi_{\theta}(y_{[a,b]}|y_a) g_{\eta}(x_{[a,b]}|y_{[a,b]})
q(y^{new}_{[a,b]}|y_a,r^{new}_{tot},\theta)} \right\}  .
\end{equation} 

\subsection{Choice of the sub intervals $[a,b]$}
\label{sub-int}
To ensure that the process can be updated on the whole interval
$[t_0,t_n]$, we have to choose a suitable set of sub intervals
$\mathcal{I}_{[t_0,t_n]}$ for which we apply the above updating
algorithms. As a general rule, one can say that they should be
overlapping. Also it is often useful to include sub intervals which do not lead
to a change of the process at the observation times $t_1<t_2<\dots<t_n$. In
such situations, the terms  $g_{\eta}(x_{[a,b]}|y^{new}_{[a,b]})$ and
$g_{\eta}(x_{[a,b]}|y_{[a,b]})$ are equal and
therefore cancel out in the acceptance probability. 

In cases where the observations are complete and noise-free,
we need only sub-intervals of the form $[t_{k-1},t_k]$. However,
because it is sometimes a non-trivial problem to find a realization
of the process which matches all observations, we found that it
is sometimes useful to include a tiny noise in the model and to
choose also sub-intervals with a $t_k$ as interior point. By this
trick we can often obtain realizations that match all observation
by the above updating algorithms. 

In general, good choices of the sub-intervals can be
very dependent on the given situation. The standard one is to let
$\mathcal{I}_{[t_0,t_n]}$ consist of all intervals of the form
 $[t_{k-1},t_k]$ and $[(t_{k-1}+t_k)/2,(t_{k}+t_{k+1})/2]$.

\subsection{Updating the path at a border} \label{updateborder}
In the cases $b=t_n$  or $a=t_0$ we also want to change the values of
$y_{t_n}$ and $y_{t_0}$, respectively (unless $f_0$ is a Dirac
measure). We recommend to propose first a change in
$r_{tot}^{new}$, that is 
\begin{equation}\label{simrtotneweb}
r_{tot}^{new} = r_{tot} +r',\ r' \sim q_{\iota'}^{r'},
\end{equation}
where $q_{\iota'}^{r'}$ is a symmetric distribution on $\mathbb{Z}^r$. Then
either $y_a$ or $y_b$ remains unchanged and the other value follows from 
$y_b-y_a=A r_{tot}^{new}$. The rest can be done
again with Algorithm \ref{propyb}. If $y^{new}_{t_0} \neq y_{t_0}$, the factor
$f_0(y^{new}_{t_0})/f_0(y_{t_0})$
is needed additionally in the acceptance probability (\ref{accprob}).
In the examples, we discuss how to proceed if we want to change only some
components of $y_{t_0}$ or $y_{t_n}$, respectively.

\section{Initialisation of $\eta$, $\theta$ and
  $y_{[t_0,t_n]}$}\label{ini}
The form of the trajectories of the underlying Markov jump process depends
strongly on the parameter $\theta$ and the value at $t_0$. So just choosing
$\eta^{(0)}$ and $\theta^{(0)}$ and then simulating
$y_{[t_0,t_n]}^{(0)}$ leads 
usually to processes which match the observed data badly. It then takes
very many iterations in the algorithm  until we obtain processes that
are compatible with the data.

In our experience, generating the starting values by algorithm 
\ref{inialgo} below leads to substantial increases in computational efficiency.
It is inspired by the particle filter: We select the most likely particle,
perform 
a number of Metropolis-Hastings steps (similarly to \cite{gwbc}) and
propagate with the Gillespie algorithm. 

An additional trick can bring further improvement. Because the speed of the
techniques described depends heavily on the number of 
reactions in the system, one wants to ensure that the initial value
$y_{[t_0,t_n]}$ for Algorithm \ref{simyhetaeta} has
rather too few than too many reactions. We can achieve this with
a simple shrinkage factor $\nu$ between $0$ and $1$ for 
$\theta$ during the initialisation, that is replacing $\theta$ after
simulation with $\nu\cdot\theta$. This
acts like a penalisation on the reaction numbers: It does not affect the
probabilities in (\ref{gill2}) (the time-homogeneous case), but makes the
system slower, resulting in fewer reactions.  

\begin{algo}[Generating starting values]
\label{inialgo}
\mbox{}
\begin{enumerate}
\item Choose $\eta^{(0)}$. 
\item Simulate $S^{\{1\}}$ i.i.d starting values 
  $y_{t_0}^s \sim p(y_{t_0}|x_{t_0})$ and generate 
  $y_{(t_{0},t_1]}^s$ for $s \in \{1,2,\dots,S^{\{1\}}\}$ using
  the Gillespie algorithm with the normalized standardized reaction
  intensities $\mathbb{I}_{\{h_i>0\}}$ ($i=1,\dots,r$) and equal
  hazard rates $1/(t_1-t_0)$. Set 
  $y^{\{1\}}_{[t_0,t_1]}=y_{[t_0,t_1]}^{s'}$, where
  $s'=\argmax_s\{g_{\eta^{(0)}}(x_{[t_0,t_1]}|y^s_{[t_0,t_1]})\}$.
  Simulate  $\theta^{\{1\}} \sim
\Gamma_r\left(\tilde{\alpha}(y^{\{1\}}_{[t_0,t_1]}),\tilde{\beta}(y^{\{1\}}_{[t_0,t_1]})\right)$.
\item For $l=1,\dots,n-1$:
\begin{itemize}
\item[a)] Use $M^{\{l\}}$ steps of algorithm \ref{simyhetaeta} on
  $[t_0,t_l]$ with
  shrinkage factor $\nu$  and starting values $y_{[t_0,t_l]}^{\{l\}}$ and
  $\theta^{\{l\}}$  to generate 
  $y_{[t_0,t_l]}^{\{l+1\}}$ and $\theta^{\{l+1\}}$.  
\item[b)] Generate $S^{\{l\}}$ paths $y_{[t_0,t_{l+1}]}^{s}$ which are 
  independent continuations of
  $y_{[t_{0},t_{l}]}^{\{l+1\}}$ on $(t_l,t_{l+1}]$, based on the Gillespie 
  algorithm with  $\theta^{\{l+1\}}$. Set
  $y^{\{l+1\}}_{[t_0,t_{l+1}]}=y_{[t_0,t_{l+1}]}^{s'}$,
  where
  $s'=\argmax_s\{g_{\eta^{(0)}}(x_{l+1}|y_{t_{l+1}}^s)\}$.  
\end{itemize}
\item Set $\theta^{(0)}=\theta^{\{n\}}$ and 
  $y^{(0)}_{[t_0,t_n]}=y^{\{n\}}_{[t_0,t_n]}$.
\end{enumerate}
\end{algo}

So to propagate to the process on the interval $(t_l,t_{l+1}]$ (for 
$l=1,\dots,n-1$), we use 
$\theta^{\{l+1\}}$ which should roughly follow the distribution of
$\theta|x_{[t_{0},t_{l}]},\eta^{(0)}$, because of step 3.a).

\section{Examples}\label{ex}
\subsection{Stochastic Oregonator}\label{oregsec}
\begin{figure}[h]
  \centering
  \includegraphics[scale=.9]{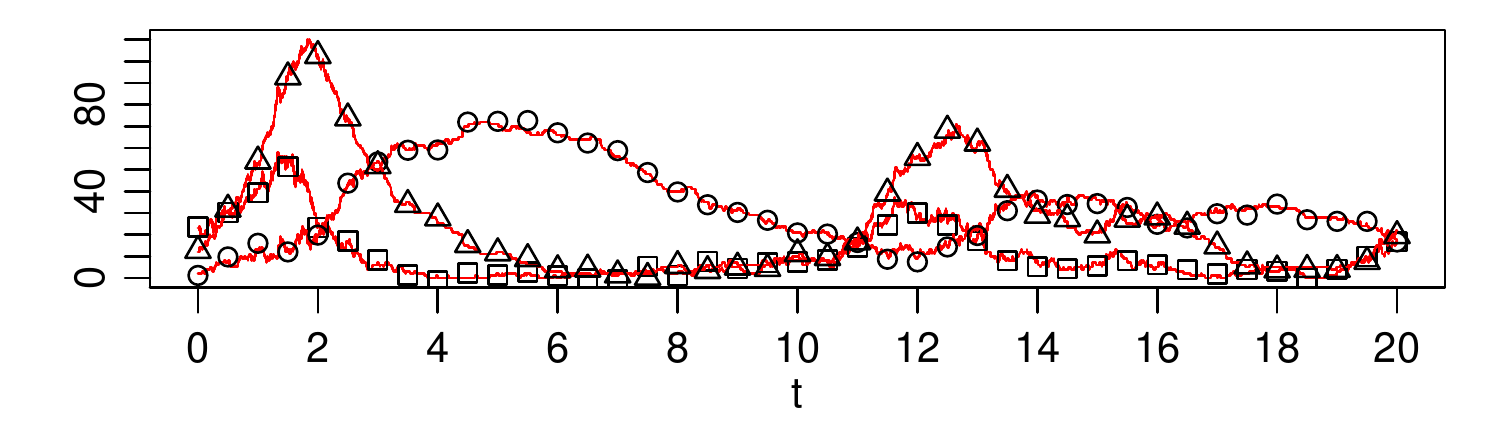}
\caption{Observations $x^1_{t}$ (squares), $x^2_{t}$ (circles) and
  $x^3_{t}$ (triangles) of the Oregonator model for $t \in
  \mathbb{T}$. The true Markov jump process is indicated as
solid line.}
  \label{oregonatordatap}
\end{figure}

First we consider the stochastic Oregonator to illustrate the
algorithms. It is a highly idealized model of the Belousov-Zhabotinskii
reactions, a non-linear chemical oscillator. It has 3 species and the
following 5 reactions:

\begin{displaymath}
  \begin{array}{llcl}
    R_1:& Y^2  &\longrightarrow& Y^1 \\
    R_2:& Y^1 + Y^2 &\longrightarrow& \emptyset \\
    R_3:& Y^1  &\longrightarrow& 2Y^1+Y^3 \\
    R_4:& 2Y^1  &\longrightarrow& \emptyset  \\
    R_5:&  Y^3 &\longrightarrow& Y^2 \\ 
  \end{array}
\end{displaymath}
For further details, see \cite{g}. Following Section \ref{model}, the
process $\{y_t:t\geq t_0\}$, where $y_t=(y^1_t,y^2_t,y^3_t)^T$ and $y^i_t$
is the number of species $Y^i$ at time $t$, is a Markov jump process
with standardized reaction intensities
\begin{equation*}
  h(y)=(y^2, y^1y^2,y^1,y^1(y^1-1)/2,y^3)^T
\end{equation*}
and the jump matrix
\begin{equation*}
  A:=\left (
    \begin{array}{ccccc}      
      1  & -1 &  1  & -2 &  0    \\
      -1  & -1 &  0  &  0 &  1    \\
      0  &  0 &  1  &  0 & -1    \\

  \end{array}
\right).
\end{equation*}
As starting distribution, we use the uniform distribution on
$\{0,\dots,K\}^3$ with $K=25$.  The measurement errors are normally
distributed with precision $\eta$, that is
\begin{equation}\label{erroreg}
  g_{\eta}(x,y)=\prod_{j:x^j\neq\na}\frac{\sqrt{\eta}}{\sqrt{2\pi}}
  \exp\left(-\frac{\eta}{2}(x^j-y^j)^2\right).
\end{equation}
In Figure \ref{oregonatordatap}, a sample trajectory for
$\theta=(0.1,0.1,2.5,0.04,1)$ and $\eta=1/2$, simulated with the Gillespie
algorithm, is shown, observed every $0.5$ units of time during a time
period of $20$.

If we choose a Gamma$(\alpha,\beta)$ prior for $\eta$, then the full
conditional distribution of $\eta$ in the posterior is again a Gamma
distribution with parameters
\begin{equation*}
  \tilde{\alpha}^{\eta}(x_{[t_0,t_n]},\alpha)=\alpha+
  \frac{1}{2}\#\{(l,j)\in \{1,\dots,n\}\times\{1,\dots,r\}:x_l^j\neq \na\}
\end{equation*} and 
\begin{equation*}
  \tilde{\beta}^{\eta}(y_{[t_0,t_n]},x_{[t_0,t_n]},\beta)=\beta
  + 
  \frac{1}{2}\sum_{(l,j):x_l^j\neq \na}\left(x^j_{l}-y^j_{t_l}\right)^2  .
\end{equation*}
This yields a simple way to perform step 3. in Algorithm \ref{simyhetaeta}.

We now want to estimate the parameters and the Markov jump process from the
observations at the discrete times $\mathbb{T}=\{0,0.5,\dots,20\}$ given in
Figure \ref{oregonatordatap} in various scenarios. The total raction
numbers for the true underlying Markov jump process are
$r_{tot}=(76,417,518,92,508)^T$.

\begin{itemize}
\item[A)] Exact observation of every species, i.e., we observe
  $\{y_{t}:t\in \mathbb{T}\}$.
\item[B)] Observation of every species with errors, i.e., we observe
  $\{x_{t}:t\in \mathbb{T}\}$.
\item[C)] Observation of species $Y_1$ and $Y_2$ with errors,
  i.e., we observe $\{(x^1_{t},x^2_{t}):t\in\mathbb{T}\}$.
\item[D)] Observation of species $Y_1$ and $Y_3$ with errors,
  i.e., we observe $\{(x^1_{t},x^3_{t}):t\in\mathbb{T}\}$.
\item[E)] Observation of species $Y_2$ and $Y_3$ with errors,
  i.e., we observe $\{(x^2_{t},x^3_{t}):t\in\mathbb{T}\}$.
\end{itemize}

\subsubsection{Specifications of the algorithm}
We specify the proposal distributions and further details in our algorithm
as follows.  A basis vector matrix is given by
\begin{equation*}V(A)=\left(
    \begin{array}{ccccc}      
      1  & -1 &  0  & 1 &  0    \\
      1  &  0 &  1  & 1 &  1    \\
    \end{array}\right)^T
\end{equation*}
and we simulate $Z$ in (\ref{simrtotnew}) with $(B_1 \bar{B}_1,B_2
\bar{B}_2)^T$, where $\textrm{P}[B_l=\pm 1]=0.5 $, $\bar{B}_l\sim
\mathrm{Bin}(2,\iota)$, $\iota=0.4$ and all random numbers are independent.
For the new reaction number at the beginning on the interval
$[t_{n-1},t_{n}]$ or at the end on the interval $[t_{n-1},t_{n}]$, we want
updates which change only one component of $y_{t_0}$ or $y_{t_n}$,
respectively, to get better acceptance. In order to do this, we need the
integer solutions to
\begin{equation*}
  A_{-j,.}
  (r_{tot}^{new}-r_{tot})=0
\end{equation*}
for $j \in \{1,2,3\}$, where $A_{-j,.}$ denotes the reaction matrix without
the $j$-th row. With the techniques from Appendix \ref{dioeq}, we find
exemplarily for $j=1$ the basis vectors $v_1=(1,-1,0,0,0)^T$,
$v_2=(0,0,0,1,0)^T$ and $v_3=(0,1,1,0,1)^T$. Because the last one is
already in the kernel of $A$, we can restrict ourselves to $v_1$ and $v_2$ for the
update, i.e., we choose one of these or their additive inverses with equal
probability and add it to the total reaction number to get the new one. We
proceed analogously for $j=2$ and $j=3$.

For the parameters $\theta$, we use $\Gamma(0.1,1)$ priors. For the
scenarios B) to E), $\eta$ is unknown. We use $\eta=10$ during the
initialisation and a $\Gamma(0,0)$ improper prior afterwards so that $\eta$
can be updated with Gamma distributions. For the initialisation (Algorithm
\ref{ini}), we use $M^{\{l\}}$ and $S^{\{l\}}$ around 100 to 200 and
slight shrinking.

We use the standard set of sub-intervals described in Subsection
\ref{sub-int}. Without further tuning, we obtained average acceptance rates
of $5\%$ - $7\%$ in all the scenarios. The running time for the 100'000
iterations of Algorithm $\ref{simyhetaeta}$ together with the
initialisation (Algorithm $\ref{ini}$), coded in the language for
statistical computing R (see \cite{rr}), was about 38 hours on one core of a 2.814 GHz
Dual-Core AMD Opteron(tm) Processor 2220 with 32'000 MB RAM. A significant
speed up is expected from coding in C.

\subsubsection{Results}

\begin{figure}[h]
  \centering
  \includegraphics[scale=.9]{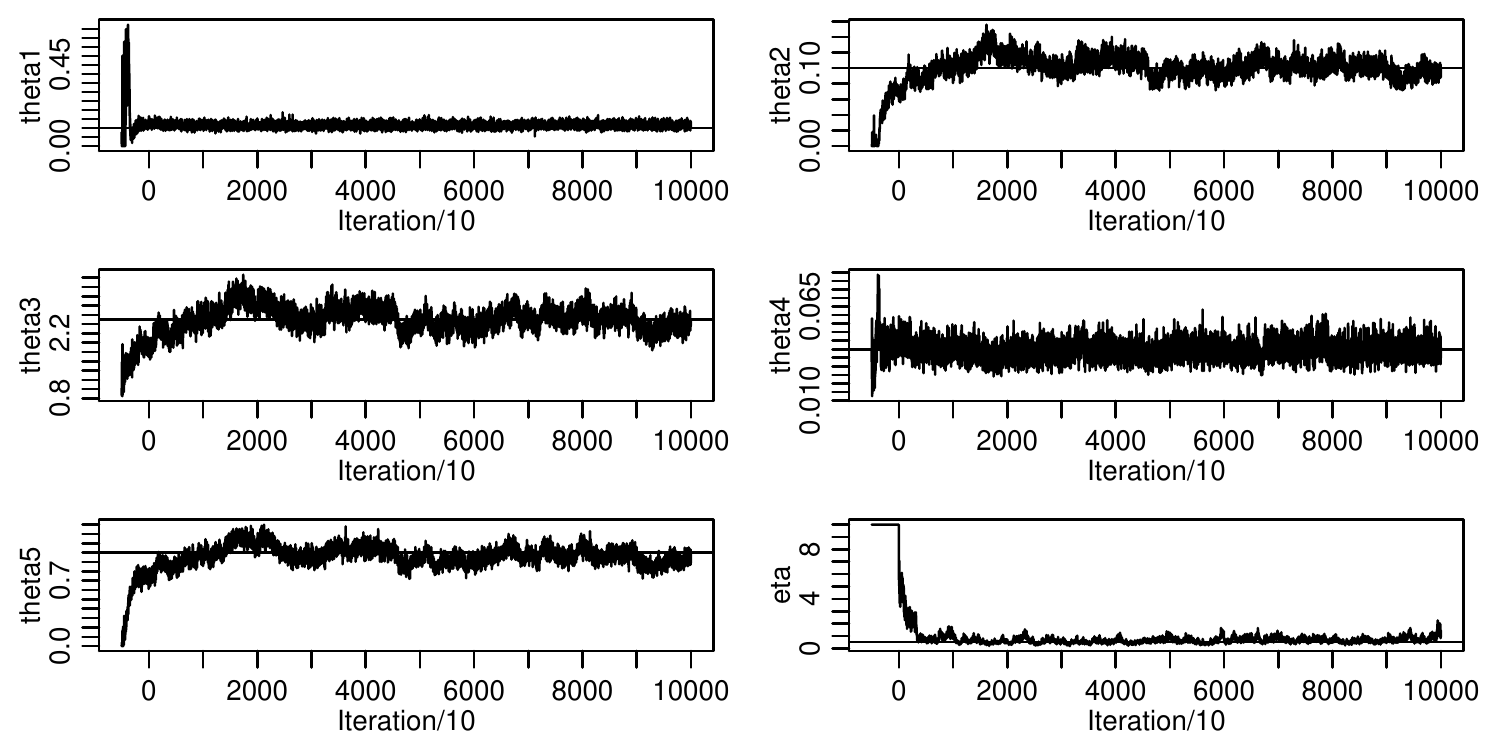}
\caption{Traces for the parameters in scenario B for the Oregonator example
  with a thinning factor 
  10. The origin on the abscissa marks the last iteration of the initialisation
  (Algorithm \ref{ini}). True
  values are indicated with a horizontal line.}
  \label{oregonatorconvp}
\end{figure}

\begin{figure}[h]
  \centering
  \includegraphics[scale=.9]{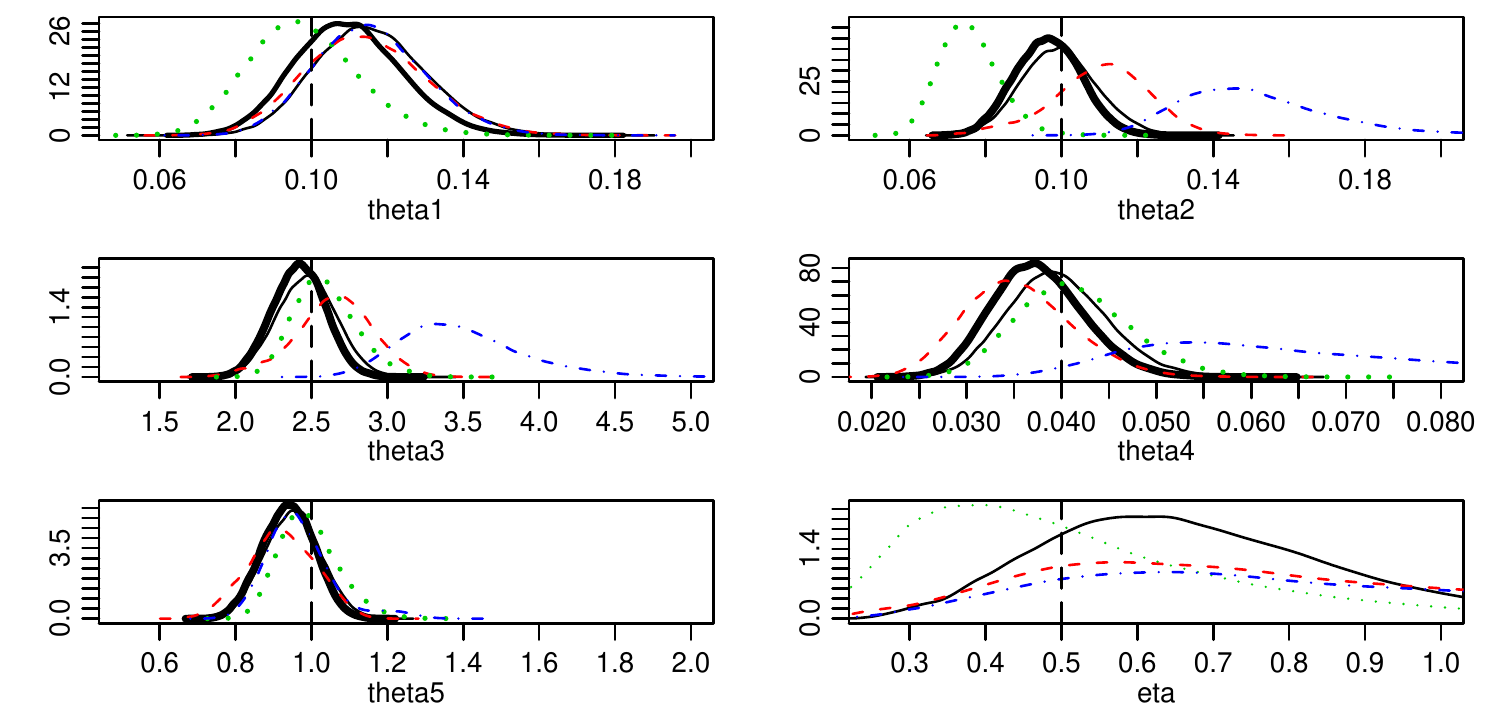}
\caption{Posterior densities of the parameters $\theta$ and $\eta$ for the
  Oregonator model in 
  the scenarios A (thick-solid), B (thin-solid), C (dashed), D (dotted) and
  E (dot-dashed), estimated
  from the last 50'000 of 100'000 iterations of Algorithm \ref{simyhetaeta}. True
  values are indicated with a vertical dotted line.}
  \label{oregonatordensp}
\end{figure}

\begin{figure}[h]
  \centering
\includegraphics[scale=.9]{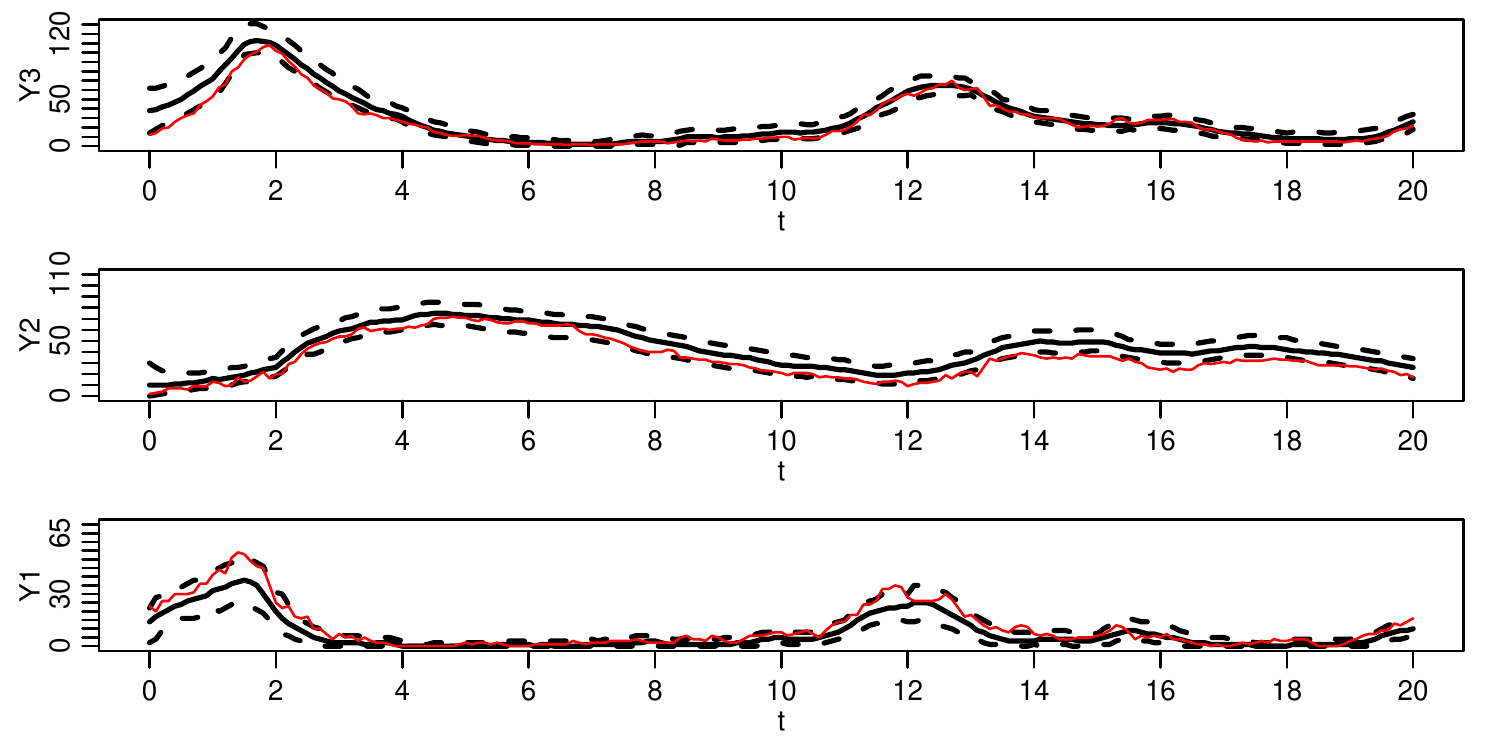}
\caption{Estimates (thick solid lines) and point-wise 95$\%$ confidence bands
  (indicated by the dashed lines) of the latent components for the
  Oregonator model
  in the scenarios C (top), D (middle) and E (bottom), respectively,
  estimated from the last 50'000 
  iterations of Algorithm \ref{simyhetaeta}, thinned by a factor of 10. The
  true values are shown as thin line.}
  \label{oregonatorestyp}
\end{figure}

In Figure \ref{oregonatorconvp}, we show the trace plots for the parameters
$(\theta,\eta)$ exemplarily in scenario B. On the whole, mixing seems
satisfactory, although not ideal for some parameters. In addition, we can
see that the initialisation process yields starting values which are
already very close to the true values.

In Figure \ref{oregonatordensp}, we compare the posterior densities
estimated from the last 50'000 of 100'000 iterations of Algorithm
\ref{simyhetaeta} in the different scenarios. The vertical dotted line
indicates the true values of $\theta$ and $\eta$, respectively. We find
that in all the scenarios the true values of $\theta$ are in regions where
the posterior density is high.  In the cases where some component is not
observable, the uncertainty is bigger, especially for the reaction rates
corresponding to standardized transition intensities which depend on this
component. For example in scenario E, $x^1$ is not observed, leading to a
loss in terms of precision for reactions rates $\theta_2$, $\theta_3$ and
$\theta_4$. The posterior densities of $\eta$ seem rather spread out, but
the mode is found nicely.

Figure \ref{oregonatorestyp} displays estimates and point-wise 95$\%$
confidence bands of the latent components in the process for the scenarios
C to E.  For comparison, we also indicate the true values of the latent
component with a thin line. We can see that they nicely lie within our
confidence bands.

\subsection{Prokaryotic auto-regulation}
We look at the simplified model for prokaryotic auto-regulation introduced
in \cite{gw} and reconsidered in \cite{gwbookart}. It is described by the
following set of 8 chemical reactions.

\begin{displaymath}
  \begin{array}{llcl}
    R_1:& \mathrm{DNA}+\mathrm{P}_2  &\longrightarrow& \mathrm{DNA}\mathrm{P}_2 \\
    R_2:& \mathrm{DNA}\mathrm{P}_2  &\longrightarrow& \mathrm{DNA}+\mathrm{P}_2 \\
    R_3:& \mathrm{DNA}  &\longrightarrow& \mathrm{DNA}+\mathrm{RNA} \\
    R_4:& \mathrm{RNA}  &\longrightarrow& \mathrm{RNA}+\mathrm{P} \\
    R_5:& 2\mathrm{P}  &\longrightarrow& \mathrm{P}_2 \\
    R_6:& \mathrm{P}_2  &\longrightarrow& 2\mathrm{P} \\
    R_7:& \mathrm{RNA}  &\longrightarrow& \emptyset  \\
    R_8:& \mathrm{P}  &\longrightarrow& \emptyset     
  \end{array}
\end{displaymath}
In this system, the sum $\mathrm{DNA}\mathrm{P}_2+\mathrm{DNA}$ remains
constant, and we assume that this constant $K$ is known and equal to 10 in
our simulation. Therefore it is enough to consider the four species
$y=(y^1,y^2,y^3,y^4)^T=(\mathrm{RNA},\mathrm{P},\mathrm{P}_2,\mathrm{DNA})^T$,
where $\mathrm{RNA}$, $\mathrm{P}$, $\mathrm{P}_2$ and $\mathrm{DNA}$ are
now interpreted as numbers of the corresponding species. According to the
mass action law, the standardized transition intensities are
\begin{equation*}
  h(y)=(\mathrm{DNA}\times\mathrm{P}_2,K-\mathrm{DNA},\mathrm{DNA},
  \mathrm{RNA},\mathrm{P}\times(\mathrm{P}-1)/2,\mathrm{P}_2,\mathrm{RNA},\mathrm{P})^T
\end{equation*}
and the jump matrix is given by
\begin{equation*}
  A:=\left (
    \begin{array}{cccccccc}      
      0  & 0 &  1  & 0 &  0  & 0  &-1  & 0   \\
      0  & 0 &  0  & 1 & -2  & 2  & 0  &-1   \\
      -1  & 1 &  0  & 0 &  1  &-1  & 0  & 0   \\
      -1  & 1 &  0  & 0 &  0  & 0  & 0  & 0   \\
    \end{array}
  \right).
\end{equation*}
As start distribution, we assume that the number of DNA molecules is
uniformly distributed on $\{0,\dots,K\}$ and the other species are
initially $0$.  Following \cite{gwbookart}, we again use normally
distributed measurement errors, see (\ref{erroreg}). The update for $\eta$
(step 4. in Algorithm \ref{simyhetaeta}), can then be done using Gamma
distributions. Also we consider three scenarios in a similar manner to the
last example.
\begin{itemize}
\item[A)] Exact observation of every species.
\item[B)] Observation of every species with errors.
\item[C)] Observation of species $\mathrm{RNA}$, $\mathrm{P}$ and
  $\mathrm{P}_2$ with errors.
\end{itemize}
True values of the parameters are
$\theta=(0.1,0.7,0.6,0.085,0.05,0.2,0.2,0.015)$, $\eta=0.5$ and we observe
the process at the integer times $0,\dots,50$. The total raction
numbers for the true underlying Markov jump process are
$r_{tot}=(192,190,122,53,116,99,117,7)^T$.

\subsubsection{Transformation of parameters}
As reported in \cite{gw} and \cite{gwbookart}, ratios of the parameters
$\theta_1/\theta_2$ and $\theta_5/\theta_6$, connected to the reversible
reaction pairs $R_1$, $R_2$ and $R_5$, $R_6$, respectively, are more
precise than the individual rates.  We found a similar behavior also for
$\theta_3/\theta_7$ and $\theta_4/\theta_8$. This is related to the fact
that adding or subtracting an equal number of the corresponding reaction
between two consecutive observation times does not change the values of the
Markov jump chain at these time points, making it rather difficult to tell
how many of these reaction events should be there from discrete
observations only. This implies also that there is a strong positive
dependence in the posterior between these pairs of parameters, slowing down
the convergence of the algorithm.

It is therefore much better to use the following reparameterization
\begin{equation*}
  \rho_1=\theta_1+\theta_2,\ \rho_3=\theta_3+\theta_7,\  
  \rho_5=\theta_4+\theta_8,\ \rho_7=\theta_5+\theta_6 
\end{equation*}
and
\begin{equation*}
  \rho_2=\frac{\theta_1}{\theta_1+\theta_2},\ \rho_4=\frac{\theta_3}{\theta_3+\theta_7},\  
  \rho_6=\frac{\theta_4}{\theta_4+\theta_8},\
  \rho_8=\frac{\theta_5}{\theta_5+\theta_6}.
\end{equation*}
We use $\Gamma(\alpha,\beta)$ priors with $\alpha=0.1$ and $\beta=1$ for
$\rho_l$ ($l=1,3,5,7$) and $\mathrm{B}(1,1)$ priors, i.e., uniform priors on
$[0,1]$, for $\rho_k$ ($k=2,4,6,8$). For updating e.g. $(\rho_{1},\rho_{2})$, we
factor the joint density of $(\rho_{1},\rho_{2})$ given $y_{[t_0,t_n]}$
as $p(\rho_1|\rho_2) p(\rho_2)$. Then $p(\rho_1|\rho_2)$ is a
$\Gamma(\alpha+r_{tot^1}+r_{tot}^2,\beta+\rho_2I_1+(1-\rho_2)I_2))$ density, and
\begin{equation*}
  p(\rho_2) \propto (\beta+\rho_2I_1+(1-\rho_2)I_2)^{-(\alpha+N_1+N_2)}\rho_2^{N_1}(1-\rho_2)^{N_2},
\end{equation*}
with $I_1=\sum_{k=1}^{n_{tot}+1}h_1(y_{k-1})\delta_k$ and
$I_2=\sum_{k=1}^{n_{tot}+1}h_2(y_{k-1})\delta_k$. The factor
$(\beta+\rho_2I_1+(1-\rho_2)I_2)^{-(\alpha+N_1+N_2)}$ can be approximated with
piecewise linear upper bounds, so we can simulate from $p(\rho_2)$ using an
adaptive accept-reject-method with mixtures of truncated Beta distributions
as proposals.

\subsubsection{Specifications of the algorithm}
The basis vector matrix is given by
\begin{equation*}V(A)=\left(
    \begin{array}{cccccccc}      
      1  & 1 &  0  & 0 &  0  & 0  & 0  & 0   \\
      0  & 0 &  0  & 0 &  1  & 1  & 0  & 0   \\
      0  & 0 &  1  & 0 &  0  & 0  & 1  & 0   \\
      0  & 0 &  0  & 1 &  0  & 0  & 0  & 1   \\
    \end{array}\right)^T
\end{equation*}
and for $q_{\iota}^Z$ we choose $q_{\iota}^Z(\pm \vec{e}_i)=0.1$ for $i
\in\{1,2,3,4\}$ and $q_{\iota}^Z(\vec{0})=0.2$ (see (\ref{simrtotnew})).

To get the new total reaction number for the update at the beginning, i.e.,
on the interval $[t_0,t_1]$, we have to respect that
$y_{t_0}^1=y_{t_0}^2=y_{t_0}^3=0$.  We therefore only want to change the
fourth component of $y_{t_0}$. So
\begin{equation*}
  A_{-4,.}
  (r_{tot}^{new}(y_{[t_0,t_1]})-r_{tot}(y_{[t_0,t_1]}))=0.
\end{equation*}
With the techniques from Appendix \ref{dioeq}, we find the same basis
vectors as in $V(A)$ plus the vector $v_5=(0,-1,0,2,1,0,0,0)^T$. So we use
(\ref{simrtotneweb}) with $q_{\iota}^{R}(\pm v_5)=0.5$.  Updating the total
reaction number at the end on the interval $[t_{n-1},t_{n}]$ is done as in
the previous example.

\subsubsection{Results}
\begin{figure}[h]
  \centering
  \includegraphics[scale=.9]{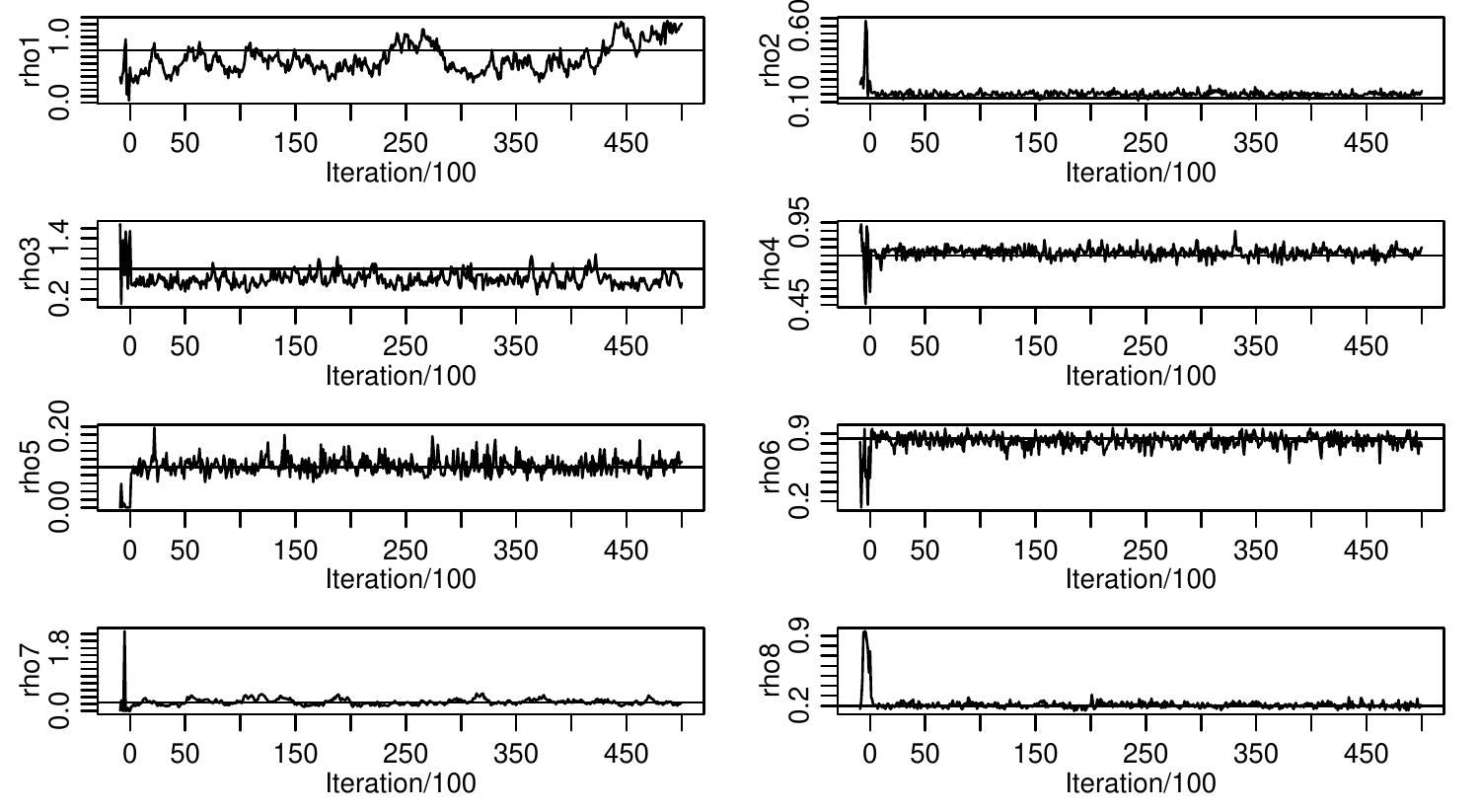}
\caption{Traces for the parameters for the prokaryotic auto-regulation
  model in scenario A with a thinning factor
  100. The origin on the abscissa marks the last iteration of the initialisation
  (Algorithm \ref{ini}). True
  values are indicated with a horizontal line.}
  \label{prokarioticconvp}
\end{figure}

\begin{figure}[h]
  \centering
  \includegraphics[scale=.9]{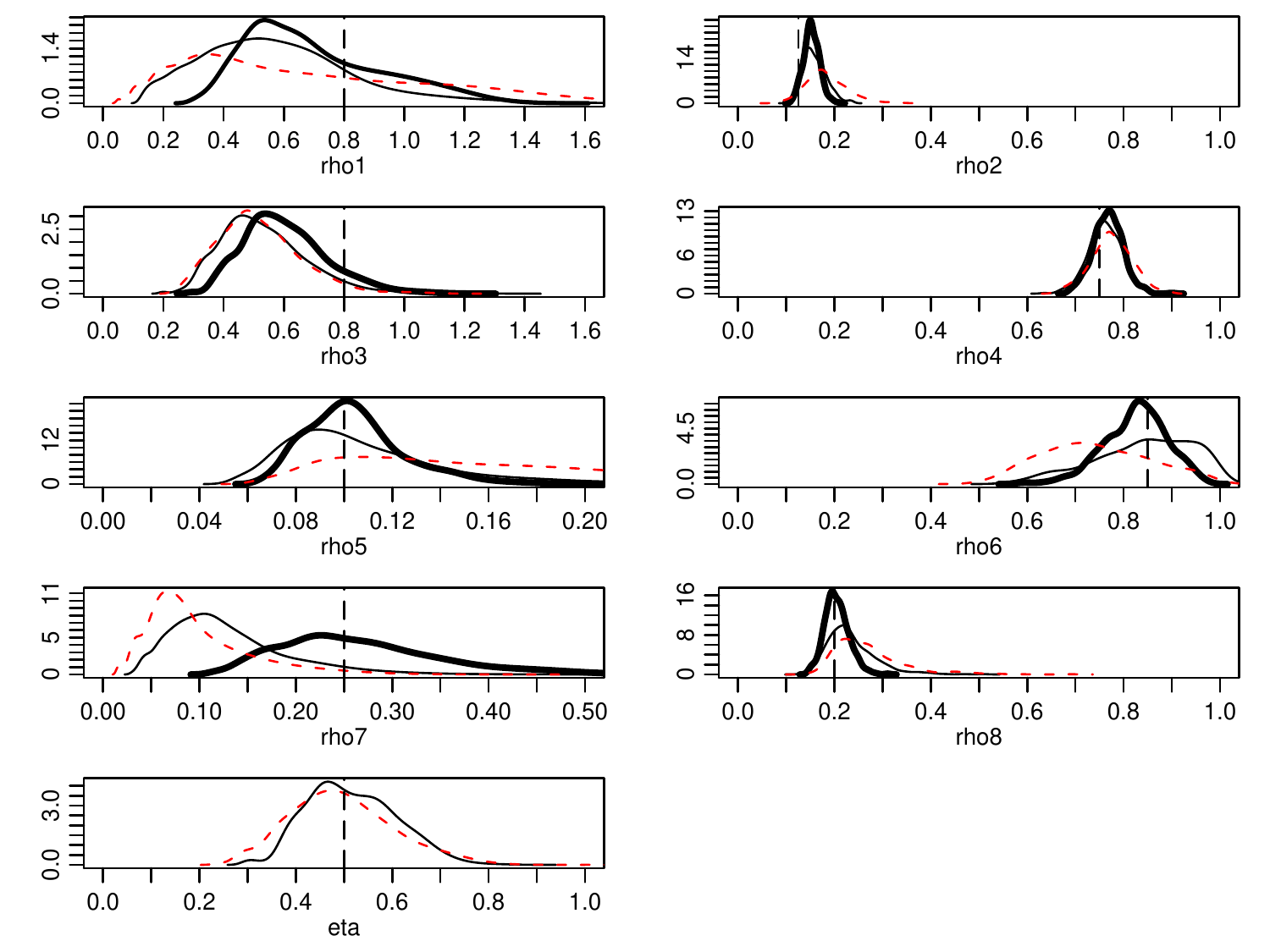}
\caption{Posterior densities of the parameters $\rho$ and $\eta$ for the
  prokaryotic auto-regulation 
  model in the scenarios A (thick-solid), B (thin-solid) and C (dashed),
  estimated 
  from the last 40'000 of 50'000 iterations (thinned with factor 100) of
  Algorithm \ref{simyhetaeta}. True values are indicated with a vertical
  dotted line.} 
  \label{prokarioticdensp}
\end{figure}

\begin{figure}[h]
  \centering
  \includegraphics[scale=.9]{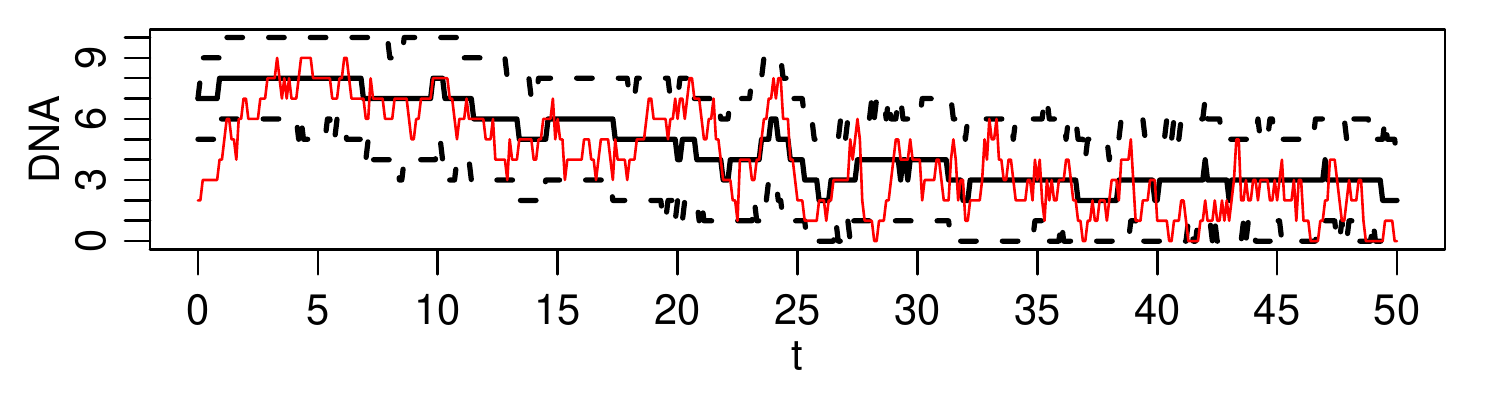}
\caption{Estimates (thick solid lines) and 95$\%$ confidence intervals
  (indicated by the dashed lines) of the latent components
  in the scenario C (prokaryotic auto-regulation 
  model), estimated from the last 40'000
  iterations of Algorithm \ref{simyhetaeta} thinned by a factor of 10. True
  values are shown as thin line.}
  \label{prokarioticestyp}
\end{figure}
The running time for the 50'000 iterations of Algorithm $\ref{simyhetaeta}$
together with the initialisation (Algorithm $\ref{ini}$), coded in the
language for statistical computing R (see \cite{rr}), was about 16 hours on on one core of
a 2.814 GHz Dual-Core AMD Opteron(tm) Processor 2220 with 32'000 MB
RAM. Average acceptance rates were around $15\%$ - $25\%$.

In Figure \ref{prokarioticconvp}, we show the trace plots of the
initialisation and 50'000 iterations of Algorithm \ref{simyhetaeta} for the
scenario A. For the parameter $\rho_1$, the mixing is somewhat problematic.

In Figure \ref{prokarioticdensp}, we can see, as expected, that the
posterior densities of $\rho_2$, $\rho_4$, $\rho_6$ and $\rho_8$ are far more
concentrated than those of $\rho_1$, $\rho_3$, $\rho_5$ and $\rho_7$. Nevertheless, all
posterior densities go well with the true values. When the number of DNA
molecules is not observed (scenario C), the posteriors of $\rho_1$ and and
$\rho_3$ are spread out, so estimating them in this scenario seems rather
hard.

Finally, we show estimate and the point-wise 95$\%$ confidence band of the
latent component in scenario C, that is the number of DNA molecules, in
Figure \ref{prokarioticestyp}. The results contain the true values quite
nicely.

\section{Conclusions}\label{concl}
In this paper, we have presented a technique to infer rate constants and
latent process components of Markov jump processes from time series data
using fully Bayesian inference and Markov chain Monte Carlo
algorithms. We have used a new proposal for the Markov jump process and - exploiting the
general state space framework - a
filter type initialisation algorithm to render the problem computationally more
tractable. Even in very data-poor scenarios in our examples, e.g. one
species is completely unobserved, we have been able to estimate parameter
values and processes and the true values are contained in posterior
confidence bands.

The techniques are generic to a certain extend, but as our examples have
shown, they have to be adapted to
the situation at hand, which makes their blind
application rather difficult. Clearly, the speed of our algorithm scales
with the number of jump events, so they are less suitable in situations
with many jumps. In such a situation, using the diffusion approximation is
recommended. However, we believe that the statement ``It seems unlikely
that fully Bayesian inferential techniques of practical value can be
developed based on the original Markov jump process formulation of
stochastic kinetic models, at least given currently available computing
hardware'' in the introduction of \cite{gwbookart} is too pessimistic.

\begin{appendix}
  \section{Integer solutions of Homogeneous Linear Equations}\label{dioeq}

  Let $A \in M_{p \times r}(\mathbb{Z})$ be an integer $p \times r$ matrix. We
  want to determine the set
  \begin{equation} \label{loesmenge} \mathbb{L}=\{x \in \mathbb{Z}^r:A
    x=0\}.
  \end{equation}
  Obviously, it is enough to consider only linear independent rows of $A$,
  so we assume $rank(A)=p\leq r$. The case $p=r$ is then trivial, so $p<
  r$.  The main idea is to transform the matrix $A$ into the so called
  Hermite normal form. For the following, see \cite{Jag}.

\begin{definition}[Hermite normal form]
  A matrix $H \in M_{p \times r}(\mathbb{Z})$ with rank $s$ is in Hermite
  normal form if

  \begin{enumerate}
  \item $\exists i_1, \dots ,i_s$ with $1 \leq i_1 < \dots <i_s \leq p$
    with $H_{i_j,j} \in \mathbb{Z} \backslash \{0\}$ for $1\leq j \leq s$.
  \item $H_{i,j}=0$ for $1\leq i \leq i_j-1,\ 1\leq j \leq s$.
  \item The columns $s+1$ to $r$ are $0$.
  \item $\lfloor H_{i_j,l}/H_{i_j,j} \rfloor=0$ for $1 \leq l <j \leq s$.
  \end{enumerate}
\end{definition}\begin{proposition}
  For every $A \in M_{p \times r}(\mathbb{Z})$ exists a unique unimodular
  matrix $U$ ($U\in GL_r(\mathbb{Z}):=\{B \in M_{r \times r}(\mathbb{Z}):
  \det(B)=\pm 1 \}$), such that $H=A U$ is in Hermite normal form.
\end{proposition}
There exist many algorithms to calculate $H$ and $U$, see e.g. \cite{Jag}.

The Hermite normal form allows us to determine the set
(\ref{loesmenge}). Because $A$ is assumed to have maximal rank, by
definition $H=(B,0)$, where $B$ is an invertible, lower triangular $p
\times p$ matrix. For $y=U^{-1} x$ we have the equation
\begin{equation*}
  0=A x= A  U  y= (B,0)  y,
\end{equation*}
so $y$ has zeroes in the first $p$ positions and arbitrary integers in the
remaining positions.  Hence a basis vector matrix $V$ for (\ref{loesmenge})
is given by $v_i=u_{r+i}$. If necessary, one can reduce the length of the
$v_i$ by the algorithm 2.3 in \cite{rip}.

\end{appendix}

\bibliography{rate-estimation-arxiv.bib}
\end{document}